# A Simulation Based Performance Evaluation of Optical Ethernet Switch

**Dawit Hadush Hailu*[1], Gebrehiwet Gebrekrstos Lema[2] and Gebremichael T. Tesfamariam[3]**

[1,2,3] School of Electrical and Computer Engineering, Ethiopian Institute of Techonology-Mekelle (EiT-M), Mekelle University, Mekelle, Tigray, Ethiopia
*Corresponding author, e-mail: dawithadush@gmail.com

*Abstract*
*With the advent of several new Cloud Radio Access Network (C-RAN) technologies, today's networking environment is dramatically altered and is experiencing a rapid transformation. One of the most important is Ethernet based C-RAN, in support of which many products such as optical Ethernet switches have recently appeared on the market. This paper presents the performance analysis of such switches with respect to Packet Loss Ratio (PLR), Latency and Packet Delay Variation (PDV). We employed the Simula based on Discrete Event Modelling on Simula (DEMOS), a context class for discrete event simulation to simulate a cut-through optical Ethernet switch under two types of traffics: High priority (HP) traffic and Low priority (LP) traffic. In this way, the paper evaluates the optical Ethernet switch performance quantitatively. The results obtained from the simulator showed that the high quality of service was reflected on HP traffic and the low quality of service in LP traffic. Hence, HP traffic can be used for transporting Radio over Ethernet (RoE) traffic while LP traffic can used for transporting time insensitive application. It is also found that HP traffic experiences a PDV equals to the duration of maximum sized LP traffic in Optical Ethernet switch.*

## 1. Introduction

Due to the ever-increasing demand for network traffic and application, the network capacity has to grow to match this demand. Newer technologies both in the optical and electrical fields allow greater capacity to cope with the huge data, but this is at the expense of the cost of the new equipment. Power consumption in cell sites and the switching devices to offer greater speed is becoming a greater and greater portion of the overall network cost. To increase the performance of the switch and reduce the power consumption in the cell sites (in mobile networks), the resources have to be used efficiently. Thus, the operators have been looking for methods to address this growing cost. As a result, a Cloud Radio Access Network (C-RAN) featuring a fronthaul network is proposed. This work basically connects the use of optical Ethernet switch in Ethernet network to the fronthaul network requirements of C-RAN.

In areas where there is dynamic user expectation like clients are engaged in mobile-savvy activities-from texting to video phone calls such as in ultra-modern stadium and dense population area (like China), requires a new emerging technology: the C-RAN. In such scenarios, using fiber as full fronthaul network may not be economically available to the rooftop where the Remote Radio Head (RRH) needs to be deployed [1]. In other cases, installing fiber in existing tower may prove to be a challenging problem. And also deploying traditional small cell sites in sub-urban and road areas (where more capacity is needed to meet fast growing traffic demand) are not realistic solutions for these areas that form a high percentage of an operator's footprint. As a complement to the both traditional small cell and fiber, a new fronthaul network that can extend the existing traditional small sites and enable quick deployment of cell site with much lower Total Cost of Ownership (TCO) is needed. Another technology such as Wavelength Division Multiplexing (WDM) and Optical Transmission Network (OTN) could save fiber consumption, however, the cost of introducing these additional transport equipment makes economically not viable for operators. Hence, the current Mobile Fronthaul (MFH) solutions are rather short-term approaches and needs improvement in both the topology and technology. As an attempt to address this issue, some of the recent research is focusing on Ethernet-based fronthaul transport network pushed by their lower costs, ability to employ statistical multiplexing, and improved performance.

So far, many researchers and scientists have studied in optical packet switched networks [2-8]. The papers focused on optimizing PLR by using different Quality of Service (QoS) differentiation policies and scheme of transport networks. The main focus of this paper is to analyzing the different QoS performance metrics that are vital in evaluating the optical Ethernet switch which will be employed in Ethernet-based fronthaul transport network. To the best of author's knowledge, the optical Ethernet switch performance is evaluated using simulation for the first time.

With packet based network realization, the performance of optical Ethernet switch issue has been one of the biggest challenges. As a result, several continuous researches have been made towards an Ethernet network for future networks. Using Ethernet in the fronthaul [9] has been proposed to take some advantages: lower cost equipment, shared use of lower-cost infrastructure with fixed access networks, obtaining statistical multiplexing, and optimized performance. Despite of their attractive advantages, Ethernet also comes with their own challenges: achieving low latency and jitter to meet delay requirements, and ultra-high bit rate requirements for transporting radio streams for multiple antennas in increased bandwidth [9]. For the above reasons, the current fronthaul networks are increasingly integrating more cost-effective packet switched technology, especially Ethernet/Internet technologies. In addition to standard Ethernet switch used by Ethernet, there is another node, H1, developed by transpacket [10] employed in IHON fusion solution. Fusion solution/IHON that uses standard Ethernet technology rather than all-optical switching technology provides the fusion properties of circuit and packet switching network in packet network. It enables Ethernet transport and ensures strict QoS for Guaranteed Service Transport (GST) traffic, and optimize resource utilization by introducing Statistical Multiplexing (SM) traffic in the unused capacity.

From this we intend to study the performance of optical Ethernet switch for C-RAN using simulation. In particular we focus on three performance metrics to evaluate the optical Ethernet switch: PLR, Latency and PDV.

2. **Reasearch Method**

Background research, conference papers, white papers, International Telecommunication Union - Telecommunication (ITU-T) and Internet Engineering Task Force (IETF) standard recommendations were used to collect the QoS of Ethernet LAN in several application areas.

To achieve the objective of this paper, the programming language chosen to construct the simulation model for the fronthaul network is Simula based on Discrete Event Modeling on Simula (DEMOS) software, a context class for discrete event simulation. Moreover, Matlab has been used for post processing of raw data's from the simulator and plotting the data's with error bars. The simulations were run 10 times by varying simulation seeds for each data points, and the results were reported with 95% confidence interval.

3. **System Model for Simulation**

To study the performance of non-preemptive optical Ethernet switch while transporting Higher Priority (HP) and Low Priority (LP) traffic, the diagram shown in Figure 1 were implemented in the simulation program. In this scenario, the Ethernet switch transmits LP packet only if no packets with HP packets is available at the input queue. When HP packet arrives while LP packets are being served, the packet will be queued till the lower priority packet has finished.

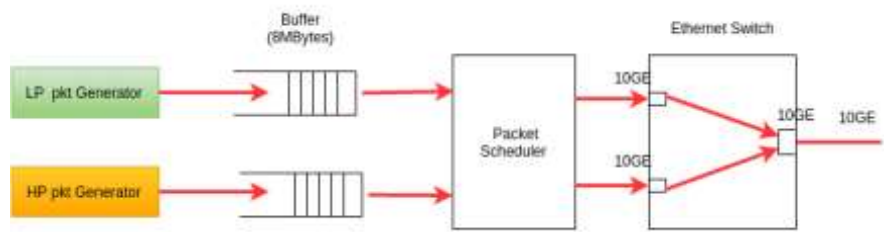

Figure 1: Illustration of Ethernet Switch for measuring the performance of HP and LP traffic

4. **Simulation Parameters**

Table I and Table II present the set of parameters which have been used during the simulation and the list of notations used for different traffic loads on different interfaces respectively.

Table I: simulation parameters used in the analysis of performance metrics of LP and HP packets.

| Parameters | Value |
|---|---|
| Seed values | 907 234 326 104 |

|  | 711 523 883 113 417 656 |
| --- | --- |
| Output link capacity | 10 Gb/sec |
| Minimum LP length | 40 Bytes |
| Length of HP packet | 1200 Bytes |
| Maximum LP length | 1500 Bytes |
| Load of HP traffic | Varies |
| Load of LP traffic | Varies |
| Buffer size | 16 MByte |
| Number of packets | 40,000 |

Table II: notation of parameters used in the simulation result analysis.

| Description | Notations |
| --- | --- |
| The load of LP traffic on 1 Gb/s interface | $L_{1GE}^{LP}$ |
| The load of HP traffic on 10Gb/s interface | $L_{10GE}^{HP}$ |
| The load of HP and LP traffic on 10 Gb/s interface | $L_{10GE}^{T}$ |

## 4. Results and Analysis

In this work, we considered PDV, PLR, and average latency as a performance metrics to evaluate performance of Ethernet switch. With analysis of each metrics, an identification of parameter that restricts the overall performance is conducted.

### 4.1. Ethernet Switch Performance

#### 4.1.1. HP Traffic Performance

In this part, the performance of HP traffic is presented with respect to: average latency, PLR, and PDV.

**Average Latency**

Figure 2 presents the average latency of HP traffic with respect to HP load for $L_{1GE}^{LP}$=0.4 and 0.45. From figure 2; we see that the average latency is increasing with increasing value of HP load. This is because HP packets have to wait in a queue when they arrive while the LP packets are serving. When the system load increases, the probability of getting the output wavelength free is low. As a result, the latency of HP packet increases. The maximum latency experienced by HP packet was 1.2 μsec.

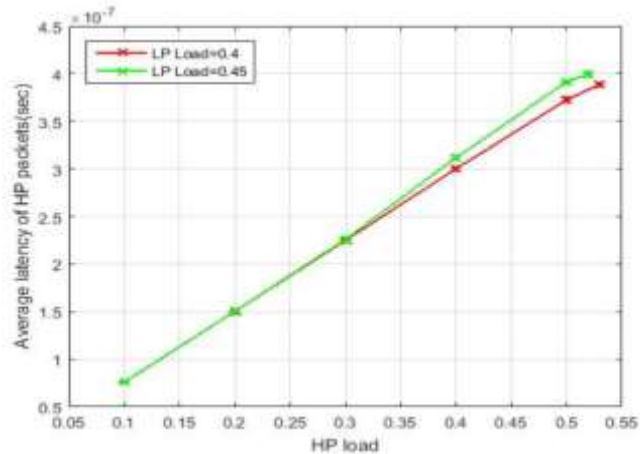

Figure 2: Average latency of HP traffic as function of HP $_{load}$ for LP $_{load}$ 0.4 and 0.45.

The average latency of HP packet is increasing with increased system load for LP $_{load}$ 0.4 and 0.45 but it has lower latency value. This is because they access the output wavelength as soon as they arrive unless the lower priority packet is serving.

**Packet Delay Variation**

Figure 3 indicates that the packet delay variation of HP traffic as a function of $HP_{load}$ for different values of $LP_{load}$. As we can see from the figure, the PDV is increased for increasing $HP_{load}$ except in the range [0.3, 0.4] and [0.5, 0.6]. However, the increment interval is not noticeable. It varies between 1.1997μsec and 1.1999μsec. This is due to the fact that the inter-packet gap for both high priority (HP traffic) and low priority (LP traffic) is not preserved which leads to the observation of this measured value. And also, the non-preemptive scheduling algorithm in Ethernet switch introduces synchronization problem, jitter [11-13]. Due to synchronization problem, PDV is going down or decreasing in the range [0.3, 0.4] and [0.5, 0.6]

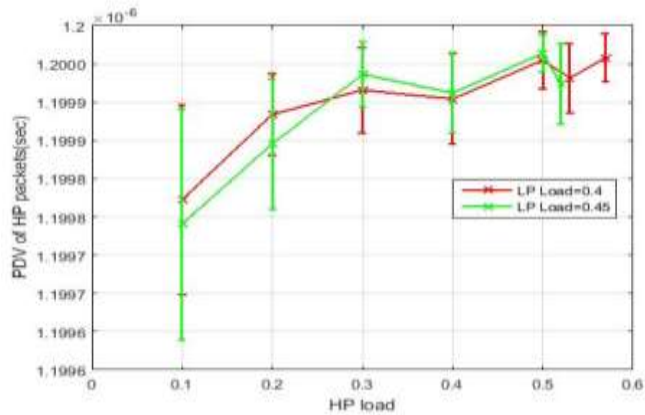

Figure 3: PDV of HP traffic as function of HP $_{load}$ for $HP_{load}$ = 0.4 and 0.45.

As discussed in Section III, in Ethernet switch, packets from HP class are always scheduled first. If the queue of this class is empty, packets from the LP class are transmitted. The HP packet experiences delay equal to the service time of the maximum length of LP packet when it arrives while the maximum LP packet is serving. Whereas, the minimum delay experienced by HP packet is zero when the HP packets are served freely. Consequently, Ethernet switch introduces PDV equals to the duration of maximum length of LP packet to HP traffic. The measured PDV value was approximately 1.2 µsec.

**Packet Loss Ratio**

In the simulation, the measured PLR of HP traffic was zero, i.e. all HP packet generated in the source were received at the output port of the Ethernet switch. This shows that the HP traffic is given higher priority than LP packets.

**4.1.2. LP Traffic Performance**

The performance of LP traffics in Ethernet switch is described below.

**Average Latency**

As shown in Figure 4, the LP packet latency increases slowly as the 10 GE HP load increases. The average Latency increases exponentially from 1 msec to 8 msec for $L_{1GE}^{LP}$=0.4 and 0.45. This is because packets with LP are transmitted only if the HP input queue has no available packet to transmit. As the HP load increases, the LP queue begins to fill up due to the arriving LP traffic while the output wavelength is not free. As a result, the amount of latency the LP packet experiences in the queue increases.

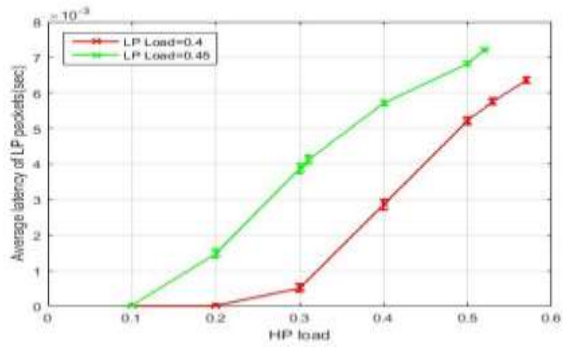
Figure 4: Average latency of LP traffic as function of HP load for LP $_{load}$=0.4 and 0.45.

The maximum latency the LP packets experience in queue is proportional to buffer size. The longer the line of LP packets in a queue waiting to be transmitted, the longer the average waiting time is.

**Packet Delay Variation**

Figure 5 presents the PDV of LP traffic as function of HP load for $L_{10GE}^{LP}$=0.4 and 0.45. Form the figure we see that the PDV increases for increasing values of $L_{10GE}^{HP}$. Here, the important observation is that the PDV for LP load $L_{10GE}^{LP}$=0.45 is higher than the PDV for LP load $L_{10GE}^{LP}$=0.4. Since LP traffic is treated as best effort traffic, the minimum and maximum latency value of LP packet is higher than the minimum and maximum latency value of HP packet.

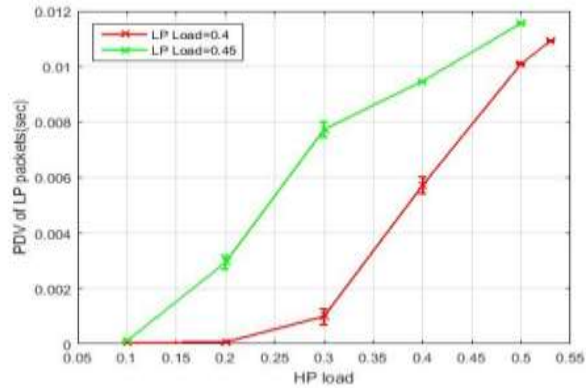
Figure 5: PDV of LP traffic as function of HP $_{load}$ for LP$_{load}$=0.4 and 0.45.

**Packet Loss Ratio**

Figure 6a shows there wasn't packet losses for the range from $L_{10GE}^{HP}$=0.1 to $L_{10GE}^{HP}$=0.45 i.e. every single LP packet sent was transmitted via the output wavelength. However, when $L_{10GE}^{HP}$ is 0.45, there was a PLR of 0.0001. Afterwards, the PLR has sharply increased when the system load is increased. Since the LP queue has a finite buffer size, at high system load the LP queue receives packets while HP traffic is serving. The increasing HP load causes the LP traffic to stay longer time at the queue. As a result, the queue of LP becomes full in short time. In this case, the LP packets arriving after the full queue are discarded and the number of discarded LP packet increases for increasing HP load. The results for $L_{10GE}^{LP}$=0.4 and 0.45 are presented in Figure 6. It is worth mentioning that increasing more traffic while the buffer size of HP and LP packets is fixed will causes buffer overflow which in turn increases the number of packets lost exponentially. The observed result proved that packets getting dropped after the buffer of the packet are full.

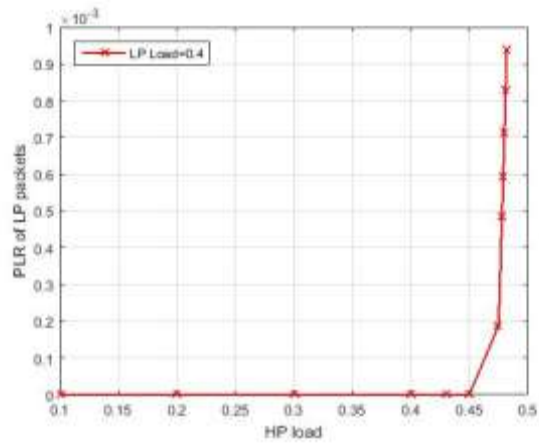

(a) $LP_{load} = 0.4$

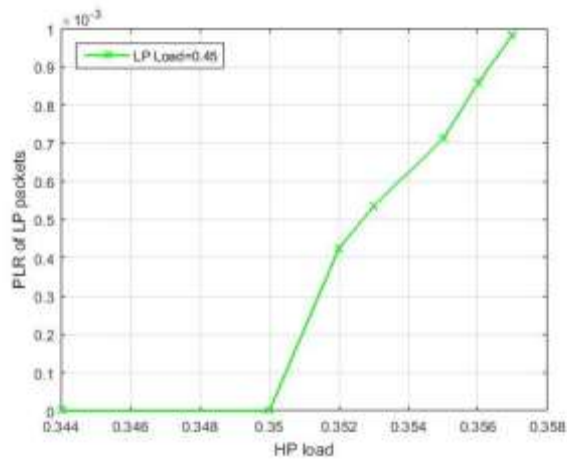

(b) $LP_{load} = 0.45$

Figure 6: PLR of LP traffic as function of HP load for $LP_{load}$ =0.4 and 0.45.

**5. Conclusion**

In this work, the overall performance of optical Ethernet switch for C-RAN network was analyzed. The high QoS was reflected on HP traffic while Low Priority (LP) traffics are not sensitive to QoS metrics and should be used for transporting time insensitive applications and services. For a scenario where there is high priority class traffic (HP traffic) and lower priority class traffic (LP) on 10 GE wavelength, the measured average latency for the higher priority class was approximately between 0.01 $\mu$sec and 0.55 $\mu$sec for $LP_{load}$=0.3, and the measured maximum latency was 1.2 $\mu$sec. The measured PLR of HP traffic in Ethernet switch was zero. Comparing the value with fronthaul requirement, Ethernet network performs better than what is recommended in IEEE 802.1CM standard. Hence, the performance of Ethernet switch fits into C-RAN fronthaul requirement in PLR comparison. As a result, the HP classes can be used for transporting Radio over Ethernet (RoE) traffic. Additionally, the measured performance metrics of HP classes are much lower than what is recommended in ITU-T recommendation for the most sensitive applications,

Acknowledgements

The authors would like to acknowledge Ethiopian Institute of Technology- Mekelle, Ethiopia, for supporting this study. Portions of this work were presented and published in thesis form in fulfillment of the requirements for the MSc. degree for one of the author's, Dawit Hadush Hailu from NTNU [14].

Conflict of interest

The authors declare that there are no conflicts of interest.

BIBLIOGRAPHY

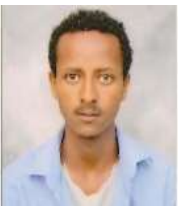
***Dawit Hadush Hailu*** received his BSc degrees in Electrical and Computer Engineering from Ethiopian Institute of Technology-Mekelle (EIT-M), Mekelle University, Mekelle, Ethiopia, in 2013 and his MSc degree from Norwegian University of Science and Technology (NTNU), Trondheim, Norway, in 2016. He is currently working as *a* Assistant Professor *in* EiT-M, Mekelle University. His research interest is in the area of networking with special emphasis on Cloud Radio Access Network (C-RAN), mobile fronthaul, radar systems, Signal Processing, Software Defined Network (SDN), optical networking and antenna design.

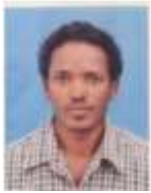
**Gebrehiwet Gebrekrstos Lema** has received his BSc in Electronics and Communication Engineering from Mekelle Institute of Technology (MIT), Mekelle University, in 2010 and his MSc in Communication Engineering from Ethiopian Institute of Technology-Mekelle (EiT-M) in 2015. He was working as a lecturer in EiT-M, Mekelle University and currently he is attending his PhD in TU of Ilmenau, German. His research interest focuses in antenna design, Self-Organized networks, cellular future networks, optimization technique, beam forming, radar systems, mobile and wireless communication, Signal Processing, Data and Computer Networking.

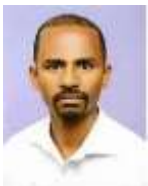
**Gebremichael T. Tesfamariam** (**Dr. –Ing.**) has got his BSc in Electrical Engineering and MSc in Control Engineering from Addis Ababa University in 2001 and 2005 respectively. He studied his PhD in Signal Processing at Technische Universität Darmstadt, Germany in 2013. Currently, he is head of School of Electrical and Computer Engineering, EiT-M, Mekelle University, Ethiopia. His research interests are Signal Processing, Radar Systems, Antenna design and Adaptive control systems.